\documentclass[aps, prb, twocolumn, superscriptaddress, showpacs]{revtex4}
\usepackage{amssymb}
\usepackage{amsmath}
\usepackage{bm}
\usepackage[dvips]{graphicx}
\usepackage{dcolumn}
\usepackage{longtable}
\usepackage{color}
\usepackage[dvipdfm]{hyperref}

\begin{document}

\title{Ferromagnetic Critical Behavior in U(Co$_{1-x}$Fe$_x$)Al ($0 \leq x \leq 0.02$) Studied by $^{59}$Co Nuclear Quadrupole Resonance Measurement}
\author{K.~Karube}
\altaffiliation{karube@scphys.kyoto-u.ac.jp}
\author{T.~Hattori}
\author{K.~Ishida}
\altaffiliation{kishida@scphys.kyoto-u.ac.jp}
\affiliation{Department of Physics, Graduate School of Science, Kyoto University, Kyoto 606-8502, Japan}
\author{N.~Kimura}
\affiliation{Department of Physics, Graduate School of Science, Tohoku University, 
Sendai 980-8578, Japan}
\affiliation{Center for Low Temperature Science, Tohoku University, Sendai 980-8578, Japan}
\date{\today}

\begin{abstract}
In order to investigate physical properties around a ferromagnetic (FM) quantum transition point and a tricritical point (TCP) in the itinerant-electron metamagnetic compound UCoAl, we have performed the $^{59}$Co nuclear quadrupole resonance (NQR) measurement for the Fe-substituted U(Co$_{1-x}$Fe$_x$)Al ($x$ = 0, 0.5, 1, and 2\%) in zero external magnetic field. 
The Fe concentration dependence of $^{59}$Co-NQR spectra at low temperatures indicates that the first-order FM transition occurs at least above $x$ = 1\%.
The magnetic fluctuations along the $c$ axis detected by the nuclear spin-spin relaxation rate $1/T_2$ exhibit an anomaly at $T_\mathrm{max}$ $\sim$ 20 K and enhance with increasing $x$. 
These results are in good agreement with theoretical predictions and indicate the presence of prominent critical fluctuations at the TCP in this system.
\end{abstract}

\maketitle
\section{Introduction}
\begin{figure}[tbp]
\begin{center}
\includegraphics[width=6cm]{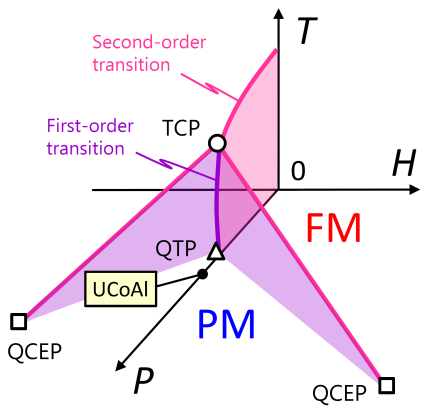}
\end{center}
\caption{(Color online) Schematic 3D [Temperature ($T$) $-$ magnetic field ($H$) $-$  pressure ($P$)] phase diagram for itinerant FM compounds. 
The pink and purple lines show the second-order and first-order transition lines, respectively. TCP, QTP, and QCEPs are denoted as a circle, a triangle, and squares, respectively. The first-order (metamagnetic) transition ``wings" spreading from the first-order transition line connecting the TCP and the QTP toward the QCEPs are shown by the purple planes. The PM ground state in UCoAl at $(T, H, P)$ = 0 is located in the vicinity of the QTP.}
\label{Fig1_3DPhaseDiagram}
\end{figure}

Itinerant ferromagnetic (FM) compounds have attracted much attention because some of them, for example, MnSi \cite{Pfleiderer}, ZrZn$_2$ \cite{Uhlarz}, and UGe$_2$ \cite{Taufour}, exhibit a similar three dimensional (3D) [temperature ($T$) - magnetic field ($H$) - pressure ($P$)] phase diagram as schematically depicted in Fig.\;\ref{Fig1_3DPhaseDiagram}. 
In this phase diagram, the FM transition temperature is suppressed by applying pressure and changes from the second-order to the first-order transition at a tricritical point (TCP) and reaches a quantum transition point (QTP) at zero temperature. 
Here we use the name QTP as a first-order quantum phase transition without criticality and it differs from a commonly used ``quantum critical point (QCP)" as a second-order quantum phase transition.
From the first-order transition line connecting the TCP and the QTP, the first-order metamagnetic transition ``wings" emerge into finite magnetic fields with a critical endpoint (CEP) of the first-order transition, and finally the CEP terminates at a quantum critical endpoint (QCEP) at zero temperature. 
This characteristic phase diagram is theoretically well described by Yamada \cite{Yamada} and Belitz \textit{et al}. \cite{Belitz}.
However, clear experimental results about criticality in the TCP, for example, quantitative estimations of critical fluctuations or critical exponents, have been insufficient, since example materials clearly exhibiting TCPs are rare and in most cases TCPs appear only under pressure of 1-2 GPa \cite{Pfleiderer,Uhlarz,Taufour}. 

\begin{figure}[tbp]
\begin{center}
\includegraphics[width=7cm]{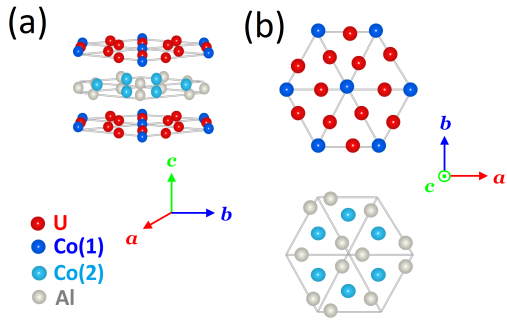}
\end{center}
\caption{(Color online) (a) ZrNiAl-type hexagonal crystal structure (space group: $P\overline{6}2m$, No. 189) of UCoAl composed by U-Co(1) layer and Co(2)-Al layer alternately stacking along the $c$ axis. 
Co(1) and Co(2) are crystallographically inequivalent atomic sites.  
(b) U-Co(1) layer and Co(2)-Al layer from the view of the $c$ axis. 
U-U bonds and Al-Al bonds form the distorted Kagom{\'e}-lattices in U-Co(1) layer and Co(2)-Al layer, respectively.}
\label{Fig2__CrystalStructure}
\end{figure}

UCoAl, we report here, is expected to be one of the itinerant FM compounds following the above 3D phase diagram.
UCoAl crystallizes in the hexagonal ZrNiAl-type structure with alternately stacked U-Co(1) and Co(2)-Al layers as shown in Fig.\;\ref{Fig2__CrystalStructure}. Note that Co(1) and Co(2) are crystallographically inequivalent atomic sites. 
At ambient pressure, its ground state is paramagnetic (PM), but it undergoes a first-order metamagnetic transition by small magnetic fields only applied along its easy magnetization axis ($c$ axis) of $\mu_0 H_{\parallel c}$ $\sim$ 0.6 T at low temperatures. The first-order transition changes to crossover above the CEP at $(T, \mu_0 H_{\parallel c})$ $\sim$ (12 K, 1.0 T) \cite{Mushnikov,Combier,Aoki,Nohara,Karube,Palacio}. 
Under hydrostatic pressure ($P_\mathrm{hydro}$), the metamagnetic transition field increases \cite{Mushnikov} and the CEP reaches the QCEP at $(T, \mu_0 H_{\parallel c}, P_\mathrm{hydro})$ $\sim$ (0 K, 7 T, 1.5 GPa) \cite{Aoki,Combier}.
In contrast, uniaxial pressure along the $c$ axis ($P_{\parallel c}$) plays a role of a ``negative pressure" and induces the FM state without magnetic field above as small as $P_{\parallel c}$ = 0.04 GPa \cite{UCoAl_uniaxial_1,UCoAl_uniaxial_2,Karube_UCoAl_uniaxial}.
The FM phase is also induced by appropriate elemental substitutions, for instance, U(Co$_{1-x}$Fe$_x$)Al exhibits the FM state at ambient pressure above as small as $x$ = 1\% \cite{UCoAl_Fedope_1,UCoAl_Fedope_2}.
These experimental facts indicate that the PM ground state in UCoAl is located in the vicinity of the QTP in the 3D phase diagram as shown in Fig.\;\ref{Fig1_3DPhaseDiagram}.
However, there has been insufficient explicit experimental evidence of the existence of the TCP in UCoAl.
Our recent $^{27}$Al nuclear magnetic resonance (NMR) measurement under uniaxial pressures pointed out that the maximum of magnetic fluctuations around ($T$, $P_{\parallel c}$) $\sim$ (20 K, 0.16 GPa) in the magnetic field along the $a$ axis ($\mu_0$$H_{\parallel c}$ = 0) is a sign of criticality in the TCP \cite{Karube_UCoAl_uniaxial}, 
but inhomogeneity of pressure makes it difficult to determine whether the FM transition is first order or second order.
In order to capture more convincing evidence of the existence of the TCP and the character of magnetic fluctuations, we have performed $^{59}$Co nuclear quadrupole resonance (NQR) measurement on the Fe-substituted system U(Co$_{1-x}$Fe$_x$)Al in zero external magnetic field. NQR is one of the most powerful techniques to provide both static and dynamic magnetic properties in zero external magnetic field.
\section{Experiment}
The polycrystalline samples U(Co$_{1-x}$Fe$_x$)Al ($x$ = 0, 0.5, 1, and 2\%) were synthesized by use of the arc melting method in a tetra-arc furnace.
The polycrystalline ingots were crushed into fine powders ($\sim$0.2 g) and packed into plastic tubes for the present NQR measurement. 
We have also performed the powder x-ray diffraction measurement to characterize synthesized samples.
The NQR measurement was performed by a typical spin-echo method irradiating a $\pi/2$ pulse and a $\pi$ pulse at a time interval of $\tau$ using a superheterodyne-type spectrometer.
\section{Results}
\begin{figure}[tbp]
\begin{center}
\includegraphics[width=7cm]{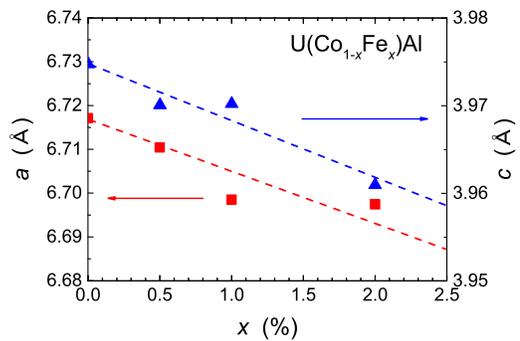}
\end{center}
\caption{(Color online) Fe concentration ($x$) dependencies of the lattice constants $a$ (squares) and $c$ (triangles) estimated by the powder x-ray diffraction peaks. 
The broken lines show the linear relationship expected from Vegard's law.}
\label{Fig3_Xray}
\end{figure}

The obtained powder x-ray diffraction patterns were consistent with the simulation without additional peaks, namely, there are no spurious phases in the samples. The calculated lattice constants $a$ and $c$ in the hexagonal structure slightly decrease by increasing Fe concentration (Vegard's law) as shown in Fig.\;\ref{Fig3_Xray}.

\begin{table}[tbp]
\caption{Nuclear quadrupole parameters of $^{59}$Co(1), $^{59}$Co(2), and $^{27}$Al nuclei in UCoAl}
\begin{center}
\begin{tabular}{c|cccc} \hline
Nucleus & $\nu_{zz}$ (MHz) & $\eta$ & Reference \\ \hline
$^{59}$Co(1) & 0.695 & 0 & Iwamoto \textit{et al}. \cite{Iwamoto} \\ 
$^{59}$Co(2) & 4.313 & 0 & Iwamoto \textit{et al}. \cite{Iwamoto} \\ 
$^{27}$Al & 0.385 & 0.327 & Karube \textit{et al}. \cite{Karube} \\ \hline
\end{tabular}
\end{center}
\label{Table_NQRparameters}
\end{table}

In order to analyze $^{59}$Co-NQR spectra, we use the following general form of the NQR Hamiltonian with the consideration of internal fields \mbox{\boldmath{$H$}}$_\mathrm{int}$,
\begin{eqnarray}
\mathcal{H}=\frac{\hbar\nu_{zz}}{6}\left\{ (3 I_{z}^{2}-\mbox{\boldmath{$I$}}^2)+\frac{\eta}{2}(I_{+}^{2}+I_{-}^{2})\right\}
-\gamma \hbar \mbox{\boldmath{$I$}} \cdot \mbox{\boldmath{$H$}}_\mathrm{int},
\end{eqnarray}
where $\mbox{\boldmath{$I$}}$ and $\gamma$ are the nuclear spin and the nuclear gyromagnetic ratio, respectively: $I$ = 7/2 and $\gamma$ = 10.03 MHz/T for $^{59}$Co nuclei.
In addition, $\nu_{zz}$ is the frequency along the principal axis of the electric field gradient (EFG) and $\eta$ is the asymmetry parameter defined as $\eta\equiv |V_{xx}-V_{yy}|/V_{zz}$, where $V_{ij}$ is the components of the EFG tensor. In UCoAl, the EFG principal axis is parallel to the $c$ axis and the values of $\nu_{zz}$ and  $\eta$ are given in the Table \ref{Table_NQRparameters} \cite{Iwamoto,Karube}.
In the present study, we focused on the $^{59}$Co(2) site where $\nu_{zz}$ is the largest and $\eta$ = 0.

\begin{figure}[tbp]
\begin{center}
\includegraphics[width=9cm]{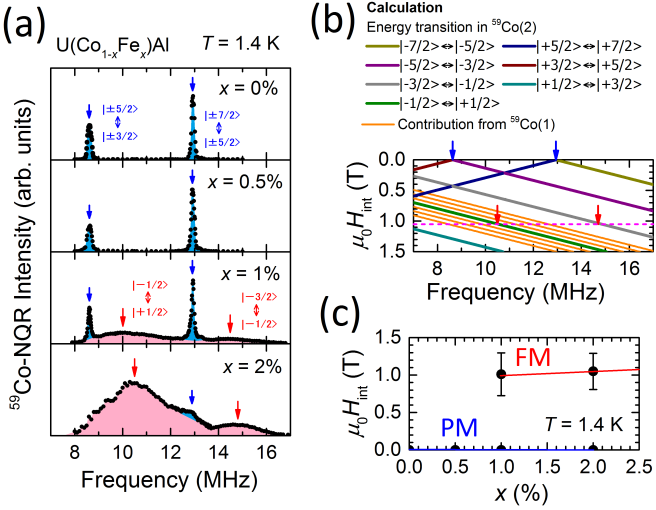}
\end{center}
\caption{(Color online) (a) $^{59}$Co-NQR spectra at $T$ = 1.4 K for $x$ = 0, 0.5, 1, and 2\%. The sharp peaks (light blue) represent the PM components and the broad peaks (pink) represent the FM components, respectively. (b) Numerical calculation of the resonance frequencies versus the internal field $H_\mathrm{int}$ along the $c$ axis. The pink broken line shows $\mu_0 H_\mathrm{int}$ = 1.05 T estimated by the values of previous studies \cite{UCoAl_Fedope_pressure_2,Nohara,Iwamoto}. (c) $x$ dependence of the experimentally estimated internal field $H_\mathrm{int}$ along the $c$ axis at $T$ = 1.4 K.}
\label{Fig4_NQRSpvsx}
\end{figure}

\begin{figure}[tbp]
\begin{center}
\includegraphics[width=8cm]{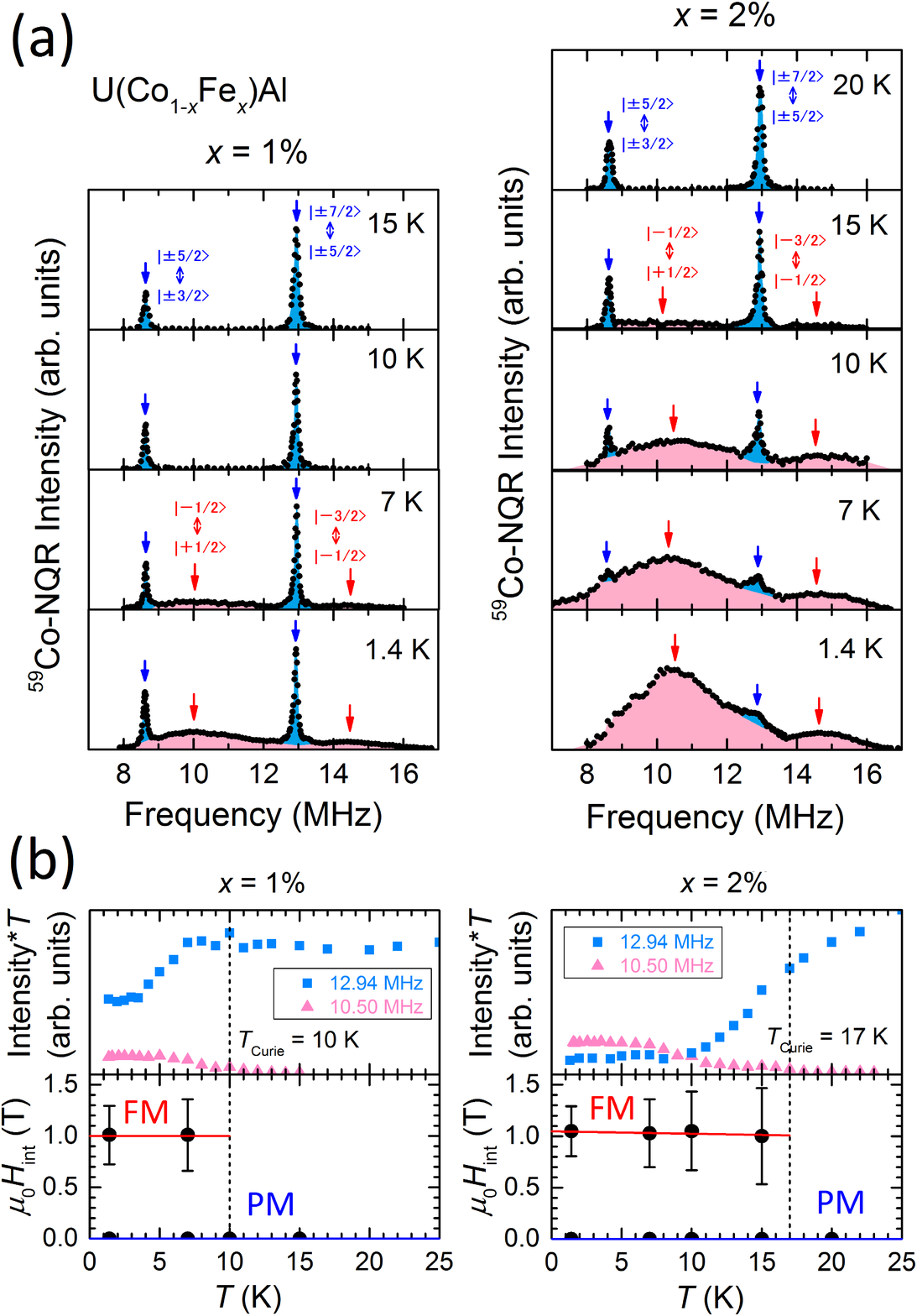}
\end{center}
\caption{(Color online) (a) Temperature dependence of $^{59}$Co-NQR spectra for $x$ = 1 and 2\%. The sharp peaks (light blue) represent the PM components and the broad peaks (pink) represent the FM components, respectively. (b) Temperature dependence of the $^{59}$Co-NQR intensity multiplied by the temperature (upper panels) and the estimated internal field $H_\mathrm{int}$ (lower panels) for $x$ = 1 and 2\%.}
\label{Fig5_NQRSpvsT}
\end{figure}

Figure\;\ref{Fig4_NQRSpvsx}(a) shows the frequency-swept $^{59}$Co-NQR spectra at $T$ = 1.4 K for $x$ = 0, 0.5, 1, and 2\%. 
In $x$ = 0 and 0.5\%, the sharp peaks were observed at $f$ = 8.63 MHz (= $2\nu_{zz}$) and 12.94 MHz (= $3\nu_{zz}$), corresponding to the transition energy of $\left| \pm 5/2 \right> \leftrightarrow \left| \pm 3/2 \right>$ and $\left| \pm 7/2 \right> \leftrightarrow \left| \pm 5/2 \right>$ in the PM state ($\mu_0 H_\mathrm{int}$ = 0), respectively. 
In $x$ = 2\%, on the other hand, the broad peaks originating from the FM components were observed around $f$ $\sim$ 10.5 MHz and 14.7 MHz, and both PM and FM signals were observed in $x$ = 1\%.
The internal field at the $^{59}$Co(2) site is parallel to the $c$ axis and its magnitude is calculated to be $\mu_0 H_\mathrm{int}$ = $A_\mathrm{hf} M_\mathrm{U}$ $\sim$ 1.05 $\pm$ 0.09 T using the results of the previous studies: the magnetic moment per U atom $M_\mathrm{U}$ = 0.37 $\mu_\mathrm{B}$ \cite{UCoAl_Fedope_pressure_2} and the hyperfine coupling constant $A_\mathrm{hf}$ = 2.58 T/$\mu_\mathrm{B}$ \cite{Nohara}, 3.07 T/$\mu_\mathrm{B}$ \cite{Iwamoto}. 
Figure\;\ref{Fig4_NQRSpvsx}(b) shows the simulation of the resonance frequencies as a function of the internal field along the $c$ axis, $H_\mathrm{int}$, calculated by diagonalizing the above NQR Hamiltonian.
With the assumption of $\mu_0 H_\mathrm{int}$ $\sim$ 1.05 T, the observed broad spectra centered at $f$ $\sim$ 10.5 MHz and 14.7 MHz turn out to arise from the transition energy of $\left| -1/2 \right> \leftrightarrow \left| 1/2 \right>$ and $\left| -3/2 \right> \leftrightarrow \left| -1/2 \right>$, respectively.
One of the reasons why the FM spectral width ($\sim$2.4 MHz) is much broader than that of the PM spectra ($\sim$0.12 MHz) is inhomogeneity of the internal field due to Fe substitution. 
The other reason is the small contribution from the $^{59}$Co(1) spectra, 
which appear around $\sim$10.5 MHz with all six quadrupole satellites within the range of 8-13 MHz as shown in the simulation (orange lines) in Fig.\;\ref{Fig4_NQRSpvsx}(b).
Note that the PM and FM spectra are superposed without continuous splitting or broadening of the PM spectra, and the intensity ratio of the FM spectra to the PM spectra increases by increasing $x$. Figure\;\ref{Fig4_NQRSpvsx}(c) shows the $x$ dependence of the internal field estimated by fitting the FM spectra in the transition energy of $\left| -1/2 \right> \leftrightarrow \left| 1/2 \right>$ with Gaussian functions. The abrupt jump of the internal field  with the phase separation suggests the first-order transition at low temperature in U(Co$_{1-x}$Fe$_x$)Al.

Figure\;\ref{Fig5_NQRSpvsT}(a) shows $^{59}$Co-NQR spectra for $x$ = 1 and 2\% at several temperatures. 
The upper panels of Fig.\;\ref{Fig5_NQRSpvsT}(b) show the temperature dependence of the $^{59}$Co signal intensity multiplied by temperature (intensity$\ast T$) at $f$ = 12.94 MHz (PM state) and 10.50 MHz (FM state) in $x$ = 1 and 2\%. 
The signal intensity is estimated from the integration of the NQR signal in a 100-kHz frequency window. 
The quantity of intensity$\ast T$ should be independent of temperature in the absence of phase transitions.
In $x$ = 1\%, the broad FM spectra appear below $T_\mathrm{Curie}^\mathrm{1st}$ = 10 K, although the peak intensity of PM spectra remains stronger than that of the FM spectra down to 1.4 K.
In $x$ = 2\%, the peak intensity of the PM spectra drops from higher temperature, and the FM spectra appear below $T_\mathrm{Curie}^\mathrm{1st}$ = 17 K and its peak intensity exceeds that of the PM spectra below 10 K. 
As discussed above, the PM and FM spectra are superposed without continuous splitting or broadening of the PM spectra. As a result, the temperature dependence of the estimated internal field shows the abrupt jump as shown in the lower panels of Fig.\;\ref{Fig5_NQRSpvsT}(b), indicating the first-order transition both in $x$ = 1 and 2\%.  

\begin{figure}[tbp]
\begin{center}
\includegraphics[width=8cm]{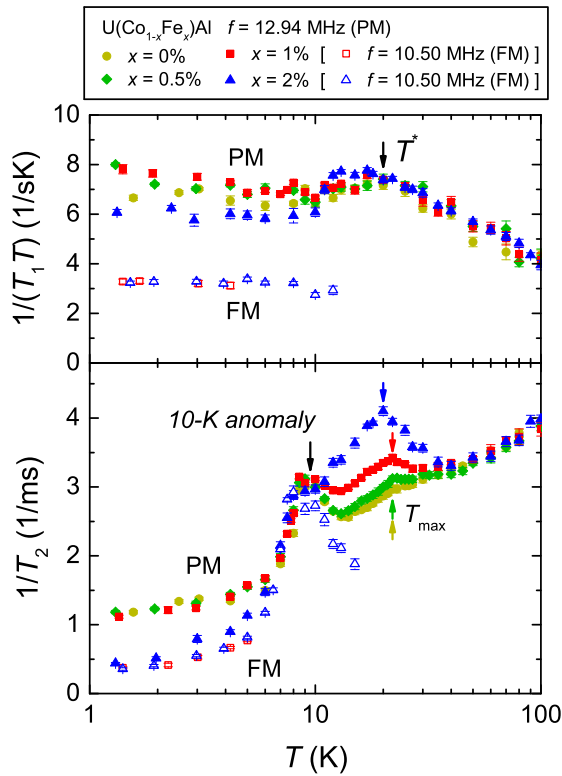}
\end{center}
\caption{(Color online) Temperature dependencies of $1/(T_1 T)$ (upper panel) and $1/T_2$ (lower panel) for $x$ = 0\% (yellow circles), 0.5\% (green diamonds), 1\% (red squares), and 2\% (blue triangles). These were measured at $f$ = 12.94 MHz (closed symbols) and $f$ = 10.50 MHz (open symbols). $T^\ast$ $\sim$ 20 K in $1/(T_1 T)$ shows the temperature below which the localized behavior changes to the itinerant behavior. $T_\mathrm{max}$, pointing to the kinks or peaks in $1/T_2$, shows the characteristic temperature where magnetic fluctuaions along the $c$ axis enhance in the PM phase. \textit{10-K anomaly} in $1/T_2$ shows the robust anomaly inherent in $T_2$ as explained in the text.}
\label{Fig6_T1T2vsT}
\end{figure}
In order to investigate the behaviors of magnetic fluctuations, 
we have measured the nuclear spin-lattice relaxation rate $1/T_1$ and the nuclear spin-spin relaxation rate $1/T_2$.
For the measurements of the spin-lattice relaxation time $T_1$, the recovery of nuclear magnetization $M$ after saturation $\pi/2$ pulses was fitted with the theoretical function for $I$ = 7/2 \cite{Narath}. 
For the measurements of the spin-spin relaxation time $T_2$, the decay of nuclear magnetization $M$ as a function of 2$\tau$ was fitted with the single exponential function $M(2\tau)=M(0)\exp(-2\tau/T_2)$. 
Although, at low temperatures (below $T$ $\sim$ 5 K), the decay curves show the compressed exponential behavior described by $M(2\tau)=M(0)\exp\left[-(2\tau/T_2)^\beta \right]$ with $\beta$ $\sim$ 1.5 due to nuclear dipolar effects, 
there is no distinct difference between the values of $T_2$ fitted by $\exp(-2\tau/T_2)$ and that fitted by $\exp\left[-(2\tau/T_2)^\beta \right]$.

The general equations of $1/T_1$ and $1/T_2$ are provided by the followings:
\begin{eqnarray}
\frac{1}{T_1}=\frac{\gamma^2}{2}\int_{-\infty}^\infty \left< \delta H_{-} (t) \delta H_{+}(0) \right> \exp(-i f t)dt, 
\end{eqnarray}
\begin{eqnarray}
\frac{1}{T_2}=\frac{1}{2T_1}+\frac{\gamma^2}{2}\int_{-\infty}^\infty \left< \delta H_{z} (t) \delta H_{z}(0) \right>dt,
\end{eqnarray}
where $\delta H_{\pm}$ [$\delta H_z$] are magnetic fluctuations perpendicular [parallel] to the quantization axis at nuclei sites.
In the present case, the quantization axis (the EFG principal axis and the internal field direction) is the $c$ axis. Therefore, $1/T_1$ and $1/T_2$ represent magnetic fluctuations in the $ab$ plane ($\delta H_{ab}$) and along the $c$ axis ($\delta H_c$), respectively.
Since UCoAl possesses Ising-type magnetic fluctuations along the $c$ axis ($\delta H_c \gg \delta H_{ab}$) \cite{Nohara,Karube}, $1/T_2$ is a more important physical quantity to detect the change of magnetic fluctuations than $1/T_1$.

The upper panel of Fig.\;\ref{Fig6_T1T2vsT} shows the temperature dependencies of $1/(T_1 T)$ in the PM component ($f$ = 12.94 MHz) and the FM component ($f$ = 10.50 MHz).
$1/(T_1 T)$ in the PM components increases on cooling (Curie-Weiss behavior) and become constant below $T^\ast$ $\sim$ 20 K (Korringa behavior) as reported by the previous NMR studies \cite{Nohara,Iwamoto}.
Clear critical behaviors were not observed in $1/(T_1 T)$ by increasing $x$. 
The small reduction below 10 K in $x$ = 2\% is caused by the contamination of the FM components because the broad FM signal overlaps at the peak of the PM signal in $x$ = 2\% below 10 K as seen in Fig.\;\ref{Fig5_NQRSpvsT}(a). 

The lower panel of Fig.\;\ref{Fig6_T1T2vsT} shows the temperature dependencies of $1/T_2$ in the PM component ($f$ = 12.94 MHz) and the FM component ($f$ = 10.50 MHz).
For $x$ = 0\%, $1/T_2$ gradually decreases on cooling and exhibits a kink at $T$ = 22 K (we call this temperature ``$T_\mathrm{max}$") and a peak at $T$ $\sim$ 10 K (we call this peak the ``10-K anomaly"). 
The kink at $T_\mathrm{max}$ corresponds to a broad maximum observed in bulk magnetic susceptibility $\chi_{H \parallel c}$ \cite{Mushnikov,Aoki,Nohara} and $^{27}$Al-NMR $1/(T_1 T)_{H \parallel ab}(\propto \left< \delta H_{c} \right>^2)$ \cite{Karube,Karube_UCoAl_uniaxial}, which is the common feature in itinerant metamagnets.
With increasing $x$ (0 $\rightarrow$ 2\%), $T_\mathrm{max}$ slightly decreases (22 $\rightarrow$ 20 K) and the magnitude of $1/T_2$ around $T_\mathrm{max}$ develops. 
This tendency is similar to the $c$-axis uniaxial pressure dependence of $^{27}$Al-NMR $1/(T_1 T)_{H \parallel ab}$ in UCoAl \cite{Karube_UCoAl_uniaxial}.
On the other hand, the 10-K anomaly is insensitive to $x$ and observed even in the FM components of $x$ = 2\%. 
The same behavior of the 10-K anomaly in $1/T_2$ was also reported in nondoped UCoAl under magnetic field up to 5 T \cite{Nohara}. 
At present, the origin of the 10-K anomaly is unclear, but we consider that the 10-K anomaly in $1/T_2$ would have nothing to do with the $c$-axis magnetic fluctuations related to the TCP, since the 10-K anomaly is independent of $x$, and no anomaly was reported in other measurements.
\begin{figure}[tbp]
\begin{center}
\includegraphics[width=9cm]{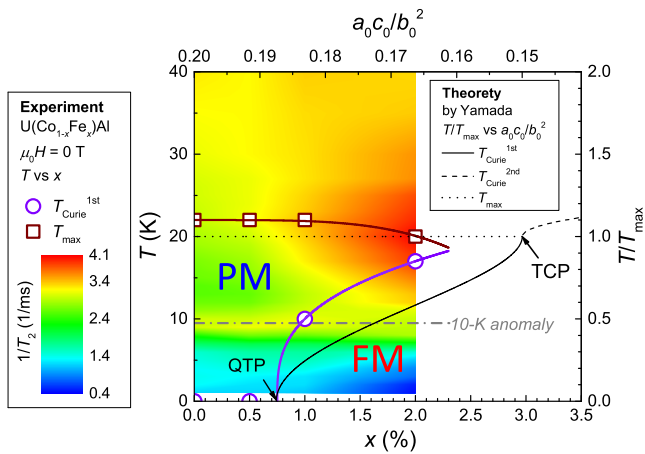}
\end{center}
\caption{(Color online) Temperature (left axis) vs Fe concentration (lower axis) phase diagram in U(Co$_{1-x}$Fe$_x$)Al and the contour plot of $1/T_2$.
$T_\mathrm{Curie}^\mathrm{1st}$ and $T_\mathrm{max}$ experimentally determined from the $^{59}$Co-NQR spectra and $1/T_2$, respectively, are denoted by open purple circles and open brown squares, respectively. 
The data points are extrapolated by eye to $x$ $>$ 2\%. 
The 10-K anomaly is denoted by the gray chain line. 
The theoretical curves by Yamada \cite{Yamada} are plotted on the $T/T_\mathrm{max}$ (right axis) vs $a_0 c_0/b_0^2$ (upper axis) phase diagram.
$T_\mathrm{Curie}^\mathrm{1st}$, $T_\mathrm{Curie}^\mathrm{2nd}$, and $T_\mathrm{max}$ are denoted by black solid, broken, and dotted lines, respectively. 
We set the scale of axes to satisfy the conditions of $a_0 c_0/b_0^2$ = 0.2 at $x$ = 0 (``pure" UCoAl), $a_0 c_0/b_0^2$ = 3/16 at $x$ = 0.75\% (QTP), and $T/T_\mathrm{max}$ = 1 at $T$ = 20 K.
}
\label{Fig7_TxPhaseDiagram}
\end{figure}

\section{Discussion}
From the above results, we constructed the $T$-$x$ phase diagram as shown in Fig.\;\ref{Fig7_TxPhaseDiagram}.
In the phase diagram, with increasing $x$, the FM first-order transition line $T_\mathrm{Curie}^\mathrm{1st}$ starts from the QTP at $(T, x)$ $\sim$ (0 K, 0.75\%) and $T_\mathrm{max}$ slightly decreases.
According to the phenomenological Landau theory, 
the ground state of itinerant FM/metamagnetic systems is determined by the tuning parameter $a_0 c_0/b_0^2$, where $a_0 (>0)$, $b_0 (<0)$, and $c_0 (>0)$ are coefficients of the Landau free energy as a function of magnetization ($M$) at zero temperature given by 
\begin{eqnarray}
F_0 (M) = \frac{a_0}{2}M^2+\frac{b_0}{4}M^4+\frac{c_0}{6}M^6.
\end{eqnarray}
Yamada \cite{Yamada} extended it to finite temperatures with introducing the thermal fluctuation term and 
showed the presence of a maximum in the temperature dependence of magnetic susceptibility $\chi(T)$ at $T_\mathrm{max}$ in the metamagnetic region.
In the case of Ising-type systems \cite{Mushnikov}, the metamagnetic region appears in the condition of $a_0 c_0/b_0^2 > (a_0 c_0/b_0^2)_\mathrm{TCP}$ = 3/20 (= 0.15).
In addition, the spontaneous FM region appears in the condition of $a_0 c_0/b_0^2 < (a_0 c_0/b_0^2)_\mathrm{QTP}$ = 3/16 (= 0.1875) and the first-order and second-order FM transition temperatures, $T_\mathrm{Curie}^\mathrm{1st}$ and $T_\mathrm{Curie}^\mathrm{2nd}$, 
are given with $T_\mathrm{max}$ and $a_0 c_0/b_0^2$ as, 
\begin{eqnarray}
T_\mathrm{Curie}^\mathrm{1st}=T_\mathrm{max}\left(1-\sqrt{\frac{80}{3}}\sqrt{\frac{a_0 c_0}{b_0^2}-\frac{3}{20}}\right)^{1/2},
\end{eqnarray}
\begin{eqnarray}
T_\mathrm{Curie}^\mathrm{2nd}=T_\mathrm{max}\left(1+\sqrt{1-\frac{20}{3}\frac{a_0 c_0}{b_0^2}}\right)^{1/2}.
\end{eqnarray}
In UCoAl, the tuning parameter is estimated as $a_0 c_0/b_0^2$ $\sim$ 0.2 at ambient pressure and it increases by applying hydrostatic pressure \cite{Mushnikov}.
In the present case, $a_0 c_0/b_0^2$ decreases from 0.2 with increasing Fe concentration.
We show the theoretical curves of $T_\mathrm{Curie}^\mathrm{1st}/T_\mathrm{max}$ and $T_\mathrm{Curie}^\mathrm{2nd}/T_\mathrm{max}$ as a function of $a_0 c_0/b_0^2$ on the same graph in  Fig.\;\ref{Fig7_TxPhaseDiagram}, 
where we set the scale to satisfy the condition of $a_0 c_0/b_0^2$ = 0.2 at $x$ = 0, $a_0 c_0/b_0^2$ = 3/16 at $x$ = 0.75\%, and $T/T_\mathrm{max}$ = 1 at $T$ = 20 K.
Although our data are insufficient to reach the TCP, the experimental phase diagram is qualitatively consistent with the theoretical one, and we can roughly estimate the TCP at $(T, x)$ $\sim$ (20 K, 2.5\%) to be a crossing point of extrapolated $T_\mathrm{Curie}^\mathrm{1st}$ and $T_\mathrm{max}$ lines.
Furthermore, we constructed the contour plot of $1/T_2$ on the $T$-$x$ phase diagram as shown in Fig.\;\ref{Fig7_TxPhaseDiagram} and 
found that the $c$-axis magnetic fluctuations at $T_\mathrm{max}$ $\sim$ 20 K drastically develop with approaching the TCP. 
This result is also consistent with the theoretical prediction \cite{Yamada} that $\chi(T_\mathrm{max})$ diverges with approaching the TCP ($a_0 c_0/b_0^2 \rightarrow 3/20+0$) as described with the following equation: 
\begin{eqnarray}
\chi(T_\mathrm{max}) = \chi(T = 0)\frac{\frac{a_0 c_0}{b_0^2}}{\frac{a_0 c_0}{b_0^2}-\frac{3}{20}}.
\end{eqnarray}
We emphasize that the ``finite temperature" critical behaviors at the TCP as observed in the present U(Co$_{1-x}$Fe$_x$)Al are characteristic for itinerant FM systems and differ substantially from the ``zero temperature" critical behaviors at the QCP as often observed in antiferromagnetic systems, such as CeCu$_{6-x}$Au$_x$ ($x$ $\sim$ 0.1) \cite{Schroder,Lohneysen}. 

\section{Conclusion}
We have performed the $^{59}$Co-NQR measurement for U(Co$_{1-x}$Fe$_x$)Al ($x$ = 0, 0.5, 1, and 2\%) in zero external magnetic field. 
In $x$ = 1 and 2\%, the NQR spectra show the coexistence of the PM and FM components without continuous shifts, indicating the first-order FM transition.
The magnetic fluctuations along the $c$ axis estimated by $1/T_2$ exhibited an anomaly at $T_\mathrm{max}$ $\sim$ 20 K and enhance with increasing $x$.
The constructed $T$-$x$ phase diagram characterized by the presence of first-order transition and $T_\mathrm{max}$ is consistent with theoretical predictions and suggests the existence of TCP around $(T, x)$ $\sim$ (20 K, 2.5\%) where magnetic fluctuations develop.

\begin{acknowledgments}
We are grateful to D. Aoki, H. Kotegawa, A. V. Andreev, S. Yonezawa, and Y. Maeno for fruitful discussions and thank T. Yamamura in IMR, Tohoku University for experimental supports. 
This work was supported by Kyoto Univ. LTM Center, the ``Heavy Electrons" Grant-in-Aid for Scientific Research on Innovative Areas  (Grant No. 20102006) from the Ministry of Education, Culture, Sports, Science, and Technology (MEXT) of Japan, a Grant-in-Aid for the Global COE Program ``The Next Generation of Physics, Spun from Universality and Emergence" from MEXT of Japan, and the Japan Society for Promotion of Science (JSPS) KAKENHI (Grant No. 23244075). One of the authors (K.K.) was financially supported by JSPS.
\end{acknowledgments}

\bibliographystyle{apsrev}

\end{document}